\documentclass[conference]{IEEEtran}
\IEEEoverridecommandlockouts
\usepackage{amsmath,amssymb,amsfonts}
\usepackage{algorithmic}
\usepackage{graphicx}
\usepackage{textcomp}
\usepackage{xcolor}
\usepackage{textalpha}
\usepackage{url}

\providecommand{\doi}[1]{\url{https://doi.org/#1}}

\begin{document}

\title{\vspace{-1.5cm} 
    \begin{center}
        \normalsize\normalfont\itshape
        This is the peer-reviewed version of the following article: Simple model of neurogliovascular unit activity with noradrenaline leading force, which has been published in final form at https://doi.org/10.1109/DCNA63495.2024.10718403. © 2024 IEEE. Personal use of this material is permitted. Permission from IEEE must be obtained for all other uses.
    \end{center}
    \vspace{0.2cm}
    \rule{\textwidth}{0.5pt} 
    \vspace{0.8cm}
    \\ 
    Simple model of neurogliovascular unit activity with noradrenaline leading force  \\
\thanks{This study was supported by the Russian Science Foundation and Ministry of Education and Science of Kursk Region, grant 24-24-20117}
}

\author{\IEEEauthorblockN{Andrey\,Yu.\,Verisokin}
\IEEEauthorblockA{\textit{Department of Theoretical Physics} \\
\textit{Kursk State University}\\
Kursk, Russia \\
ffalconn@mail.ru}
\and
\IEEEauthorblockN{Darya\,V.\,Verveyko}
\IEEEauthorblockA{\textit{Department of Theoretical Physics} \\
\textit{Kursk State University}\\
Kursk, Russia \\
allegroform@mail.ru}
\and
\IEEEauthorblockN{Alexey\,R.\,Brazhe}
\IEEEauthorblockA{\textit{Department of Biophysics} \\
\textit{Moscow State University}\\
Moscow, Russia \\
brazhe@gmail.com}
\and\IEEEauthorblockN{Stanislav\,V.\,Lagosha}
\IEEEauthorblockA{\textit{Department of Biophysics} \\
\textit{Moscow State University}\\
Moscow, Russia \\
superstas-s@mail.ru}
}

\maketitle

\begin{abstract}
In this work we present a model of a neuroglivascular unit associated with a neuronal ensemble. We combine calcium dynamics in the astrocyte, the cascade of synthesis of arachidonic acid and its derivatives, as well as the activity of the network of inhibitory and excitatory neurons with changes in local blood flow. We add equations to the model for extracellular neurotransmitters such as GABA, glutamate, and norepinephrine. We show that the resulting model can exhibit different types of behavior, and also shows a significant connection between the functioning of the neural network and its supporting components.
\end{abstract}

\begin{IEEEkeywords}
astrocytes, neuron activity, vascular tone, arachidonic acid, neurogliovascular unit, neural tissue, neurovascular models, noradrenalin.
\end{IEEEkeywords}

\section{Introduction}
The functioning of brain tissue is a complex process that involves the interaction of different types of cells and tissue structures, and is also based on different physical principles.
There are many different approaches to recording brain activity, differing both in the research method and in the scale of the area under study.
At the same time, a large number of studies are based on the direct relationship between neural activity and blood flow in a specific area of the brain~\cite{horwitz2000neural}. Additionally, methods with the introduction of microelectrodes for observing the electrical activity generated by neurons, as well as methods for visually recording the activity of astrocytes as direct participants in information processes~\cite{sagindykov2019preprocessing} have received great development.

However, an important part of the research relates to the theoretical approach, in particular the numerical modeling of processes using systems of differential equations.
Recently, modeling of a brain region has become popular within the framework of the concept of a neurogliovascular unit - NGVU~\cite{muoio2014neurovascular}, which considers the relationship between neuronal activity, the dynamically changing state of astrocytes, described through modeling of intracellular calcium dynamics, as well as the adjacent a blood vessel supplying oxygen and energy resources to cells. 

An important participant in NGVU is also the extracellular environment, into which neurotransmitters can enter as a result of leakage from the synaptic cleft~\cite{sykova2008diffusion}, gliotransmitters are purposefully secreted or brought into the bloodstream, and the concentrations of main ions can also dynamically change, directly affecting the functioning of cells through thermodynamic characteristics of molecular machines~\cite{boddum2016astrocytic}.

A reliable description and scrupulous numerical modeling of the above-mentioned processes can serve as a starting point for determining the role of specific signaling pathways in the functioning of NGVU. In this work, we used a full description of the neuronal ensemble using the ab initio Hodgkin-Huxley formalism with the addition of an extended model of astrocytic calcium danimics~\cite{verisokin2023computational} with the introduction of vascular dynamics associated with blood flow.
We also introduce into the model under consideration a calcium-dependent pathway for the synthesis of arachidonic acid and its derivatives, which are signaling molecules leading to vasoconstriction/vasodilation. 

Thus, we relate the dynamic state of astrocytes to the radius of the blood vessel. In this context, we focus on the mathematical modeling of NGVU, taking into account the norepinephrine effect on the state of nervous tissue. Norepinephrine is involved in the control of the brain tissue activity, determining the pathways of transition between wakefulness and sleep modes of individual brain regions.

\section{Modeling}
The model proposed in the work is an extended model of the classical NGVU, which combines: a blood vessel supplying oxygen with a changing radius, an astrocyte with simulated intracellular calcium dynamics, producing vasomodulatory metabolites and maintaining the homeostasis of extracellular neurotransmitters, as well as a neuronal ensemble consisting of inhibitory and excitatory neurons, modeling different modes of activity.

\begin{figure}[!htbp]
\centerline{\includegraphics[width=11.5cm]{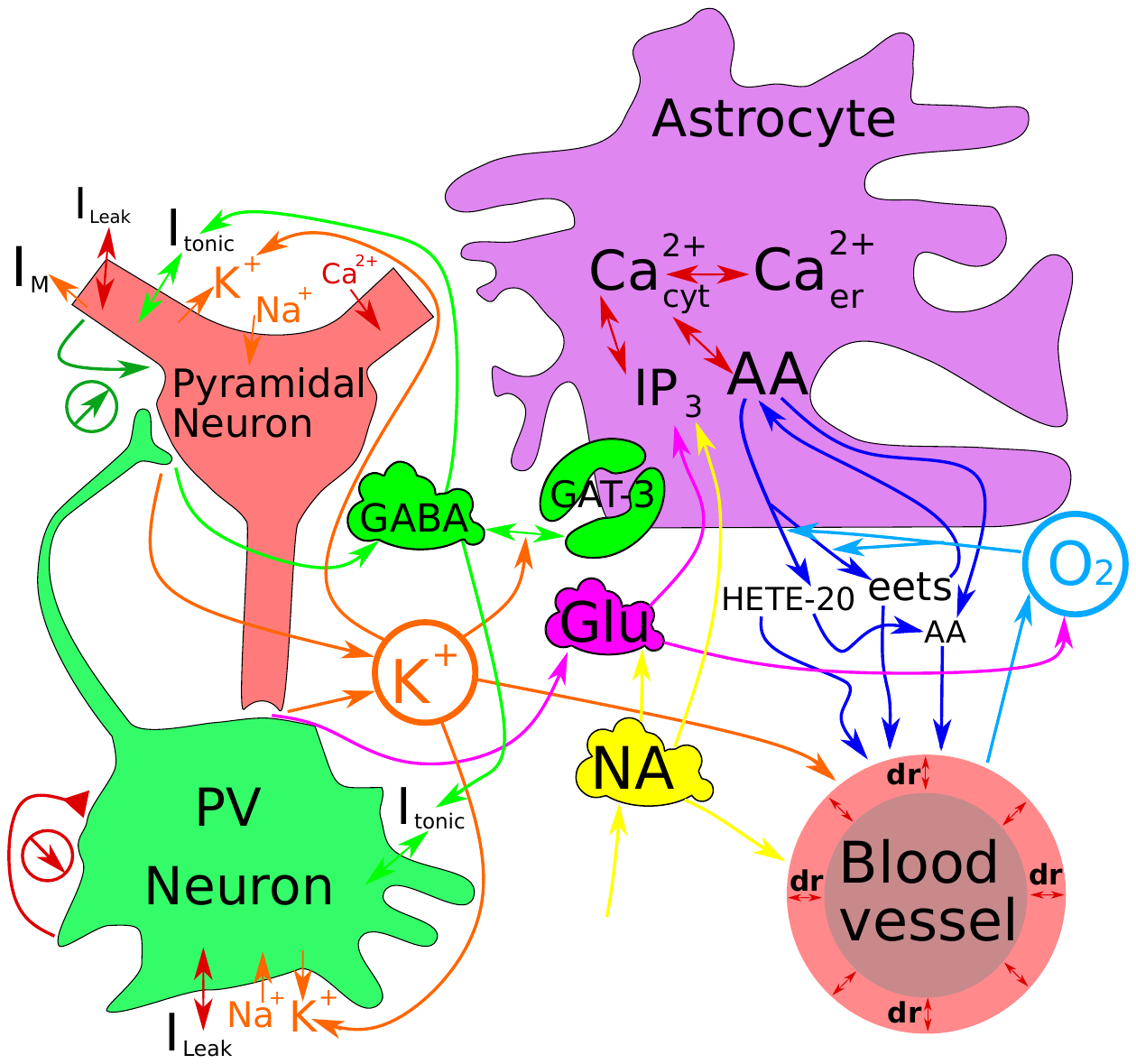}}
\caption{Schematic diagram of the interaction of NGVU elements. The directions and types of ionic currents are shown for neurons. The direction of the arrow indicates the entry of the corresponding variable into the differential equation for another variable. The paths of individual substances are color coded}
\label{FuncSheme}
\end{figure}

The functional diagram of the principal interaction paths of the proposed model is presented in Fig.~\ref{FuncSheme}.

\subsection{Neuronal part}
To adequately parameterize neurons, we will consider a region of a well-studied structure --- the hippocampus, taking as a basis layer of excitatory pyramidal cells, as well as the parvalbumin-positive GABAergic neurons associated with them. Parameters related to network connectivity, as well as synapse and excitability parameters were taken from the available experimental studies~\cite{wheeler2015hippocampome}. To maintain a balance of contributions from various cells to the model, we will consider a system of 50 excitatory and 10 inhibitory neurons per 1 astrocyte and 1 vessel. We need to somehow motivate the size of the system.

To describe the dynamics of the membrane potential of an individual cell, we used the modified Hodgkin\,--\,Huxley formalism, with ready-made parameterization based on intracellular ion concentrations and dynamic characteristics of membrane currents~\cite{aussel2022cell}. 

Transmembrane currents of excitatory pyramidal neurons are described by the following equations:
\begin{eqnarray}
\frac{{dV}}{{dt}} = {\textstyle{{( - {I_M} - {I_{leak}} - {I_{Na}} - {I_K} - {I_{Ca}} - {I_{tonic}} - {I_{EC}} - \sum {{I_{syn}}} )} \over {{C_{mem}}{S_{pyr}}}}};
\end{eqnarray}

\[{I_M} = {\sigma _{Mpyr}}{S_{pyr}}(V - {V_{revK}});\]
\[{I_{leak}} = {\sigma _{leakpyr}}{S_{pyr}}(V - {V_{revleak}});\]
\[{I_K} = {\sigma _{Kpyr}}{S_{pyr}}{n^4}(V - {V_{revK}};\]
\[{I_{Na}} = {\sigma _{Napyr}}{S_{pyr}}{m^3}{h^1}(V - {V_{revNa}});\]
\[{I_{Ca}} = {\sigma _{Capyr}}{S_{pyr}}{m_{Ca}}^2{h_{Ca}}(V - {V_{revCa}}).\]

Extracellular potassium is a connecting element between the blocks of the model that describe astrocytic and neuronal dynamics. It dynamically participates in determining neuronal excitability through ab initio principles of activity generation. Equations describing the currents of inhibitory neurons:

\begin{eqnarray}
\frac{{dV}}{{dt}} = {\textstyle{{( - {I_{leak}} - {I_{Na}} - {I_K} - {I_{tonic}} - \sum {{I_{syn}}} )} \over {{C_{mem}}{S_{pyr}}}}},
\end{eqnarray}

\[{I_{leak}} = {\sigma _{leakpv}}{S_{pv}}(V - {V_{revleak}}),\]
\[{I_K} = {\sigma _{Kpv}}{S_{pv}}{n^4}(V - {V_{revK}}),\]
\[{V_{revK}} = \frac{{RT}}{F}\log (\frac{{{K_{ext}}}}{{{K_{ipv}}}}),\]
\[{I_{Na}} = {\sigma _{Napv}}{S_{pv}}{m^3}{h^1}(V - {V_{revNa}}).\]
All activation variables n, h, m, for each type of neurons and type of currents obey general equations:
\begin{eqnarray}
\frac{{dn}}{{dt}} = {\alpha _n}(1 - n) - {\beta _n}n
\end{eqnarray}
Synapses are also organized between neurons, described by a biexponential model:
\[{I_{syn}} = {g_{syb}}(V - {V_{syn}})\sum\limits_i {{s_i}},\]
\begin{eqnarray}
\frac{{d{s_0}}}{{dt}} =  - \frac{{({s_0} - {s_\infty })}}{{{\tau _s}}},\\
\frac{{d{s_1}}}{{dt}} =  - {s_1}\beta. 
\end{eqnarray}
During an action potential, the following values of synaptic variables are set:
\[{s_0} = {s_1},\]
\[{s_1} = ({s_\infty }(1 - {e^{\frac{{ - 1}}{{{\tau _s}}}}}) + {s_1}{e^{\frac{{ - 1}}{{{\tau _s}}}}}{e^\beta }).\]

Also, when an AP arrives at any synapse, the release of potassium into the extracellular environment is taken into account, as a result of the work of potassium channels. Glutamate, released at each action potential of pyramidal cells, works in the same way.

\subsection{Astrocyte}
We adapt the model describing astrocytic dynamics, with minor changes, from an earlier work (\cite{verisokin2023computational}), which in turn is divided into several blocks.
Calcium dynamics block:
\begin{eqnarray}
\frac{{d{{[{\rm{C}}{{\rm{a}}^{2 + }}]}_c}}}{{dt}} = (1 - SVR){J_{ER}} + SVR({J_{in}} - {J_{out}});
\end{eqnarray}
\begin{eqnarray}
\frac{{d[I{P_3}]}}{{dt}} = SVR({J_\delta } + {J_{Glu}} + {J_{NA}}) - \frac{{[I{P_3}] - {{[I{P_3}]}_\infty }}}{{{\tau _r}}};
\end{eqnarray}
\begin{eqnarray}
\frac{{d{{[{\rm{C}}{{\rm{a}}^{2 + }}]}_{ER}}}}{{dt}} = (1 - SVR){J_{ER}};
\end{eqnarray}
\begin{eqnarray}
\frac{{dh}}{{dt}} = \frac{{{h_\infty } - h}}{{{\tau _h}}}.
\end{eqnarray}

Glutamate block in communication with neuronal activity:
\begin{eqnarray}
\frac{{d[Glu]}}{{dt}}{\rm{ =  }}\frac{{{{[Glu]}_0} - [Glu]}}{{{\tau _{Glu}}}} + \bar v\left( {{{[K]}_o}} \right){\rm{.}}
\end{eqnarray}

\subsection{Blood Vessel}
For simplicity, we consider blood flow to be proportional to the radius of the blood vessel, depending on the concentration of extracellular potassium, as well as vasomodulatory metabolites:
\begin{eqnarray}
\frac{{dr}}{{dt}}{\rm{ =  }}\frac{{{r_0} - r}}{{{\tau _r}}}. 
\end{eqnarray}
Oxygen concentration proportional to the radius:
\begin{eqnarray}
\frac{{dP{O_2}}}{{dt}}{\rm{ =  }}\frac{{PO_2^0 - P{O_2}}}{{{\tau _{P{O_2}}}}} + {k_{P{O_2},\,r}}r - CMR{O_2} \cdot P{O_2}.
\end{eqnarray}
\subsection{Norepinephrine}
We consider the influence of norepinephrine (NA) on both astrocyte calcium dynamics and NGVU functioning in general, through the influence of NA on [IP$_3$] production.
\begin{eqnarray}
\frac{{d[NA]}}{{dt}} = {J_{NA}} + \frac{{{{[NA]}_{amb}} - [NA]}}{{{\tau _{NA}}}} + N{A_{diff}};
\end{eqnarray}
\[{J_{NA}} = {r_{NA}} + {r_{NG}}([Glu] - {[Glu]_{amb}});\]
\[{I_{NA}} = {v_{NA}}\frac{{{{[NA]}^{0.3}}}}{{{{[NA]}^{0.3}} + k_{NA}^{0.3}}}.\]


\begin{figure}[!htbp]
\centerline{\includegraphics[width=8.5cm]{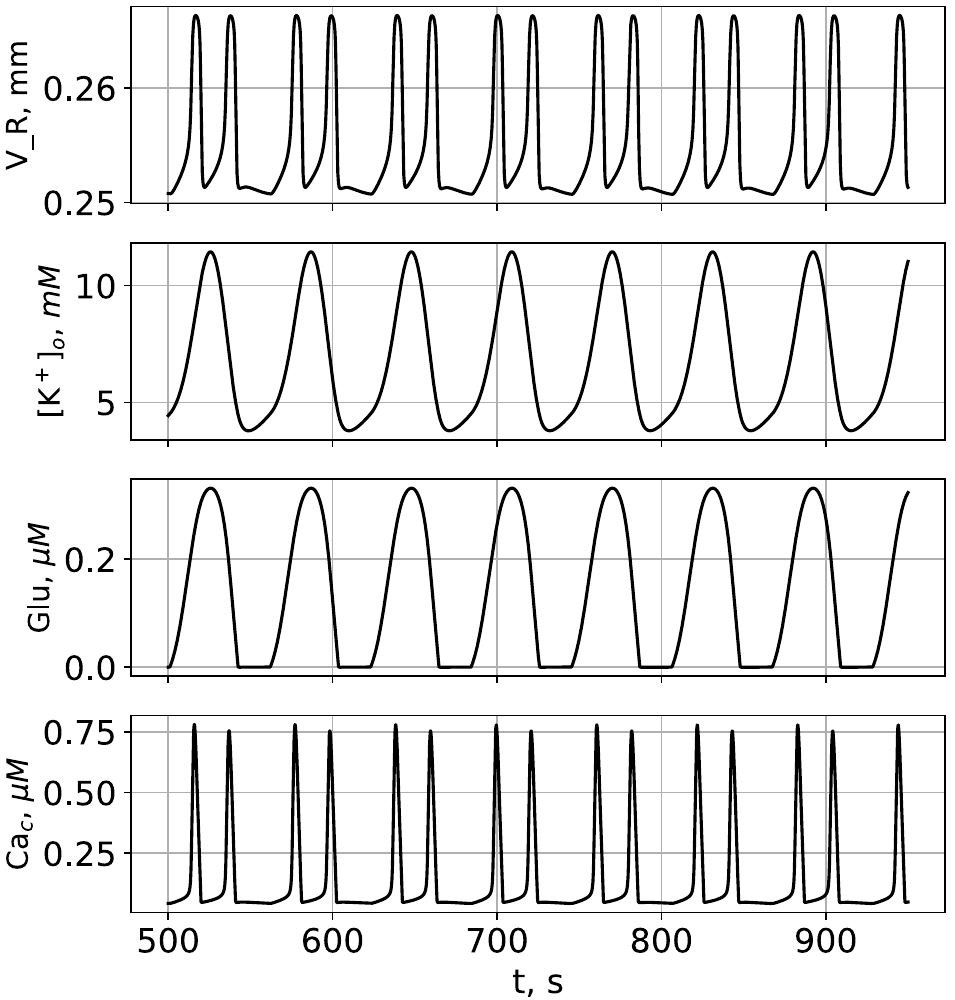}}
\caption{An example of long-term activity of NGVU. From top to bottom: changes in the radius of a blood vessel; dynamics of extracellular potassium concentration; dynamics of extracellular glutamate concentration; dynamics of intracellular calcium concentration in an astrocyte. It is worth noting, firstly, the periodic nature of the dynamics, and secondly, the doubled response of calcium concentration and vessel radius to each peak of glutamate and potassium.}
\label{SimpleSolution}
\end{figure}

\section{Numerical solution}
\begin{figure}[!htbp]
\centerline{\includegraphics[width=8.5cm]{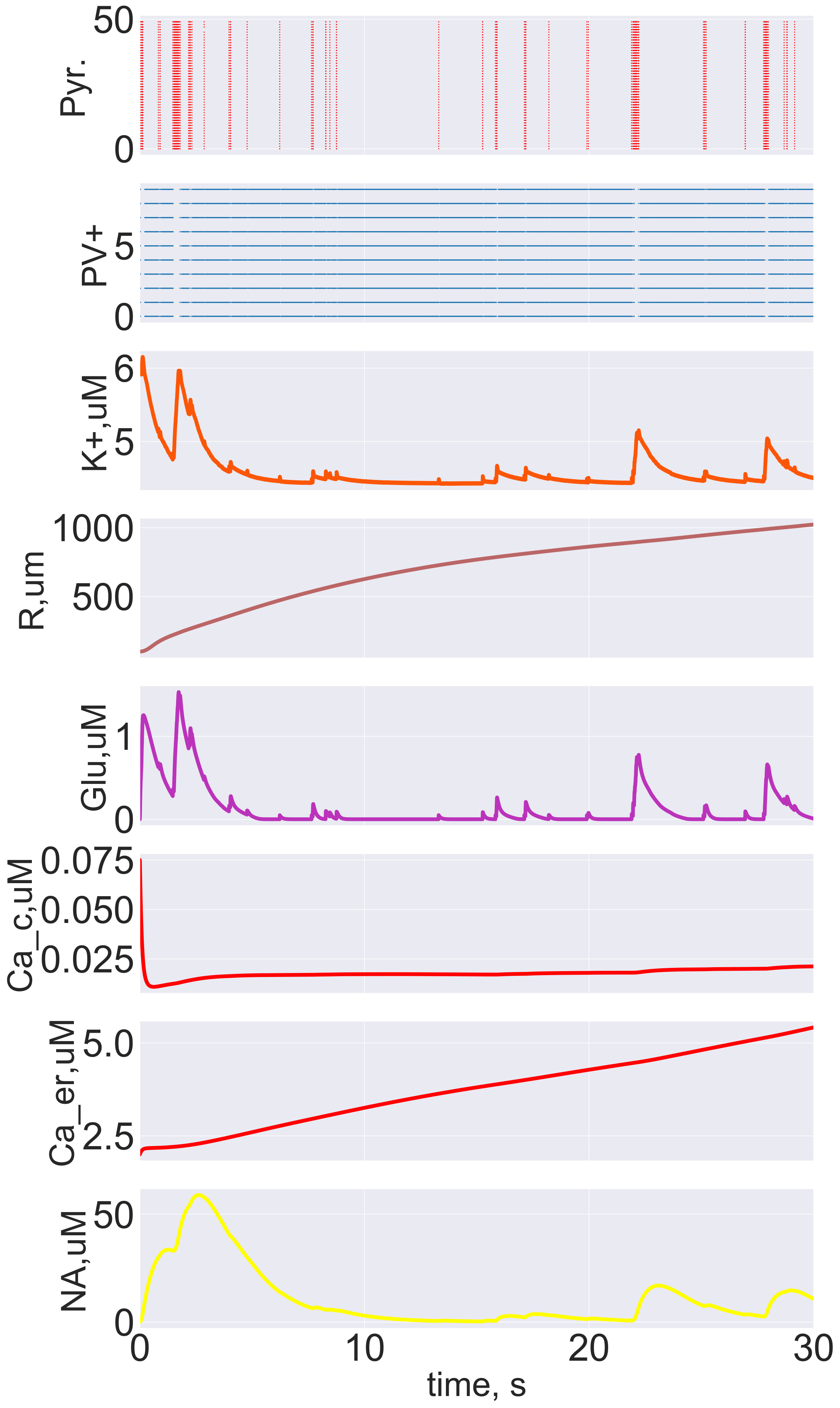}}
\caption{An example of a numerical solution of the model. From top to bottom: action potentials of pyramidal neurons and action potentials of parvalbumin positive interneurons (red and blue dots, respectively, numbers on y-axis indicates the neuron index); dynamics of extracellular potassium concentration (orange curve); change in the radius of a blood vessel (brown curve); dynamics of extracellular glutamate concentration (purple curve); dynamics of intracellular calcium concentration in an astrocyte (top red curve); dynamics of calcium concentration in an astrocyte in the endoplasmic reticulum (lower red curve); dynamics of extracellular concentration of norepinephrine (yellow curve).It is important to note the direct connection between the activity of pyramidal neurons, as the most numerous in the model, with the dynamics of glutamate, potassium and noadrenaline. Moreover, differences in the characteristic times for glutamate and norepinephrine are clearly visible. }
\label{DynamoPrimer}
\end{figure}

Numerical solutions were carried out using the Brian 2 software package~\cite{stimberg2019brian}.
Fig ~\ref{SimpleSolution} shows the long-term dynamics of the model, without the addition of neuronal ensemble and equations for norepinephrine. Fig
~\ref{DynamoPrimer} shows the solution for neuronal ensembles and changes in calcium in the astrocyte and changes in the radius of the vessel

\section{Discussion}
 It is important to note that the oscillatory dynamics of the model in Fig. ~\ref{SimpleSolution}
 and pseudostochastic behavior in the Fig. ~\ref{DynamoPrimer} have different time scales. Adding a distributed network of neurons to the model, as well as extracellular neurotransmitter variables, resulted in multiple types of activity being superimposed on each other. The behavior of variables obtained in this way can be subjected to further analysis to identify non-obvious interactions. Also, reducing the model sampling time required to add Hodgkin-Huxley dynamics allows the results to be compared with experimental observations of NGVE dynamics obtained using instruments with high time resolution.

\section{Conclusions}
The mathematical model of NGVU proposed in the work describes the complex interaction of neurons, astrocytes, blood vessels and the intercellular environment. We are closer to a complete description of the entire volume of nervous tissue necessary and sufficient for a comprehensive description of its functioning.

However, we omit the entire variety of weakly expressed ion transporters, body-wide mediators and some of the local action mediators and their receptors. Rarely represented cell types and long-term changes in metabolism through expression are not described.
It is important to note that such a detailed description seems to us to be a promising direction for the development of the model, but to implement this plan it is also necessary to develop both instrumental methods for studying the brain and the material base of computing power to support models from a large number of equations.

Nevertheless, we have included fundamentally all the main relationships that exist between the indicated elements of the NGVU, which in the future will allow us to study the processes occurring at different hierarchies of time with the required accuracy.

It is assumed that the proposed model will allow to propose potential ways to control pathological processes that arise in various disorders of the normal functioning of the brain, within the framework of the described relationships. Thus, the model has great potential both in terms of computational development and in terms of a predictive basis for new experimental work.

\section*{Acknowledgment}
This study was supported by the Russian Science Foundation and Ministry of Education and Science of Kursk Region, grant 24-24-20117.


\end{document}